\documentclass[10pt,letterpaper]{article}
\usepackage{opex3}
\usepackage{graphicx}
\usepackage{amssymb}

\newcommand\vect[1]{\mathbf{#1}}

\begin{document}

\title{Laser gate: multi-MeV electron acceleration 
and zeptosecond e-bunching}

\author{A. E. Kaplan and A. L. Pokrovsky $^*$}
\address{Electr. and Comp. Engineering Dept., 
The Johns Hopkins University, Baltimore, MD 21218}
\email {alexander.kaplan@jhu.edu}
\address{$^*$ Current address: KLA-Tencor, 1 Technology Drive, Milpitas, CA 95035}

\begin{abstract}
Relativistically-intense laser beam with large field gradient 
("laser gate") enables 
strong inelastic scattering of electrons crossing the beam.
This process allows for multi-MeV electron net acceleration 
per pass within the wavelength space.
Inelastic scattering even in low-gradient laser 
field may also induce extremely tight temporal 
focusing and electron bunch formation down to quantum, zepto-second limit. 
\end{abstract}


\section{Introduction}
In an undulator, free-electron laser, gyrotron, 
Cherenkov and resonant transition radiations, etc., 
as well as in many proposed laser accelerators, 
an electron (E) beam interacts with an electromagnetic (EM) wave
or a laser (L) beam by propagating almost parallel to it 
(i.e. almost normally to the EM-fields), 
while the energy exchange is usually facilitated by a 
small electric field component \emph{parallel} to the E-beam. 
Diffraction of the L-beam, scattering of electrons 
by the strong transverse electric field,
and the Guoy phase shift impose limitation on most 
of the "parallel" acceleration configurations. 
Alternatively, when the electrons cross the EM-wave \emph{normally} to its propagation,
the energy exchange is in most cases inhibited by ``elastic'' nature of the dominant gradient, or ponderomotive force 
\cite{cit1,cit1a,cit1b,cit1c,cit2,cit2a}.
On the other hand, recent plasma-wakefield accelerators 
\cite{cit3} demonstrated acceleration gradients 
of about 0.27 GeV/cm \cite{cit4}
presenting thus a strong challenge to 
the free-space laser acceleration techniques.
Plasma-wakefield approach was also proposed \cite{cit5}
to develop a femtosecond electron bunching \emph{via}
excitation of dynamic bistability of the nonlinear plasma wave.

In this Letter we show, however, that when a laser 
and E-beam run \emph{normally} to each other, 
two major factors acting \emph{together} 
allow for strongly \emph{inelastic} scattering and huge energy 
exchange between the beams in free space: (i) large field gradients (e.g. extremely tight laser focusing) 
allow electrons to gain and retain energy at near-$\lambda$  distance,
where $\lambda$ is the laser wavelength; 
(ii) currently available laser intensities
(up to $10^{12}$ V/cm), greatly exceeding 
a relativistic scale, $E_{rel} = k m c^2/e $, 
where $k$ is wave number and $m$ is the rest mass of electron, 
allow for energy transfer beyond $10$ MeV
and acceleration gradients two orders of 
magnitude higher than those of 
plasma-wakefield accelerators \cite{cit3,cit4}.
The relativistic intensities mitigate 
limitations on field gradient, 
while it is virtually impossible to attain 
strong non-elasticity with non-relativistic intensities. 

With EM-field polarization parallel 
to the velocity of electron, 
an electron undergoes a direct acceleration/de-acceleration 
by that field acting as a "laser gate".
Even if the E-beam passes through the gate 
without \emph{net} energy gain, 
its laser-modulated momentum can cause strong temporal, 
klystron-like focusing of the exiting E-beam 
resulting in ultra-short E-bunches with 
the quantum-limited length down to zepto-seconds.
Such a system has a potential to operate as 
a full-switch laser gate for electrons, 
a new base element of a free-electron laser or 
laser electron accelerators, generator of powerful 
ultra-short EM-pulses and E-bunches, as well as high harmonics.
The kinetic energy of E-beam required to form super-short 
E-bunches is not prohibitively high and allows one to use 
e.g. electron microscope guns for experiments.

\section{Inelastic electron scattering 
and acceleration by field ultra-gradient}
In general, the relativistic Lorentz force driving an electron in laser field, 
\begin{equation}
d \vect{p}/dt=e(\vect{E}+\vect{p} \times \vect{H}/\gamma), 
\label{eq1}
\end{equation}
where $\vect{p}$ is the electron momentum, 
and $\gamma=\sqrt{1+(p/mc)^2}$ - relativistic factor, 
depends on both electric, $\vect{E}$,
and magnetic, $\vect{H}$ fields \cite{cit6}.

If $p/mc > 1$, the $\vect{H}$-field becomes a major player, 
which may e.g. result in the sign reversal of ponderomotive
force \cite{cit2}. 
However, in the anti-node plane of an infinite standing wave, 
where an $\vect{E}$-field
peaks, the $\vect{H}$-field vanishes. 
This remains true also for an anti-node 
located precisely at the focal plane
of a \emph{focused} L-beam, where the 
phase front is ideally flat whether 
it is a paraxial Gaussian or focused to an ultimate 
$\sim  \lambda$ spot L-beam \cite{cit6} (see below Eq. (4)). 

The basic configuration chosen by us here, 
is a tightly focused linearly polarized standing wave
(propagating in the $y$-axis), 
and an E-beam that passes in 
its focal anti-node plane in the x-axis,
parallel to polarization of electric field 
$\vect{E} = \hat{e}_x F(x) \sin{(\omega t + \phi)} $ 
with phase $\phi$ and amplitude spatial profile 
$F(x)$, $F(x) \rightarrow 0$ as 
$|x| \rightarrow \infty $ \cite{cit2,cit2a}
Using normalized variables, 
$\xi = kx, \ \   \tau = \omega t, \ \  \rho =p/mc, \ \ f( \xi )=F/E_{rel}$,
we write Eq. (1) without H-term as:
\begin{equation}
d \rho/d \tau=f(\xi) \sin{(\tau+\phi)}; \ \ \ d \xi/d \tau= {\rho}/\gamma.
\label{eq2}
\end{equation} 
(Outside the anti-node we use exact solution of the 
free-space wave equation and full
Lorentz equation, see below, the 2-nd paragraph preceding Eq. (6)).
While in general even (\ref{eq2}) is not solved analytically, 
it can be well approximated in some limits.
The most common low-gradient case allows for 
adiabatic approximation resulting in a cycle-averaged ponderomotive 
force acting on a particle \cite{cit1,cit1a,cit1b,cit1c,cit2,cit2a}. 
In general, the result of this approximation is that at the exit,
(i) the particle momentum and energy do not depend on the laser phase; 
(ii) the system exhibits elastic scattering, 
i.e. electrons do not gain (or lose)  energy from (or to) the EM-field.

The fast relativistic oscillations of $\rho$ and $\gamma$ 
are comparable with $\tilde\rho$ and $\tilde\gamma$,
where sign $\tilde{}$ denotes the time-averaging. 
However, the amplitude $f_{mx}$ of fast oscillations of 
$\xi$ is small $f_{mx} \ll 1$ in nonrelativistic case, and limited by
$\pi$ - in strongly-relativistic case. 
Characterizing the field gradient by the scale 
of transverse field inhomogeneity, 
$ \xi_L $ (or L-beam amplitude spot-size at FWHM, 
$w = \lambda {\rm O} ( \xi_L )$), 
along the electron trajectory, the universal 
condition of the applicability of adiabatic, elastic scattering of 
electrons by the laser is then \cite{cit2,cit2a}:
\begin{equation}
\mu \equiv 2 \pi f_{mx}/(\xi_L \sqrt{\pi^2 +4 f_{mx}^2 }) \ll 1
\label{eq3}
\end{equation}
where $\mu$ is the major parameter of the problem. 
The system then is \emph{ergodic}, i.e. the time-averaged electron 
momentum and energy coincide with the 
averaging over all possible phases - or ensemble of electrons 
(designated by $\langle \rangle$),  $ \tilde\rho = \langle \rho \rangle$, $ \tilde\gamma = \langle \gamma \rangle$.

Non-adiabatic, inelastic scattering occurs 
when condition (3) is broken. 
It is easier to achieve $\mu \sim 1$ 
in relativistic field, $f_{mx} > 1$, 
by having $\xi_L \lesssim \pi $ ($w \lesssim \lambda $), 
than in nonrelativistic one.
In the latter case, one needs to have 
$ \xi_L \sim 2 f_{mx} \ll 1$ ($w \ll \lambda $).
(This may be stringent - but not impossible - 
condition to accomplish by using non-radiative fields due to
material structures, see below.) 
The ultimate laser gate focusing 
\emph{in free space} has $w \approx 0.8 \lambda$, 
whereby the EM-field distribution is essentially 
a fundamental transverse mode of a spherical resonator 
\cite{cit6} with the in-focal-plane profile, Fig. 1(a): 
\begin{equation}
f_{ult}(\xi) = 3 f_{mx} \left(\sin\xi - \xi\cos\xi \right) / \xi^3.
\label{eq4}
\end{equation}
The onset of inelasticity occurs when incident 
electron momentum approaches the point of 
switching from reflection to transmission modes, 
where the electron motion is sensitive to slight 
change of the phase of the laser gate
at the moment of electron entrance, 
similarly e.g. to a pendulum that comes almost to rest at the 
upper, unstable equilibrium point. 
This results in the ``phase-dispersion'' of exiting electrons; 
the system is not ergodic anymore. 
For the elastic scattering we have 
$\langle \gamma_{out} \rangle = \tilde\gamma_{out} = \gamma_0 $. 
Thus, to quantify the effect of non-ergodicity 
and inelasticity of the process, one may evaluate the net gain/loss of the electron energy at the exit from the laser gate;
$\Delta \gamma = \langle \gamma_{out} \rangle - \gamma_0 $, 
by averaging over all the phases
of randomly arriving electrons in the case of incoherent E-beam 
(preformed E-bunches should be treated separately).
\begin{figure}[h]
\begin{center}
\includegraphics[width=8.6cm]{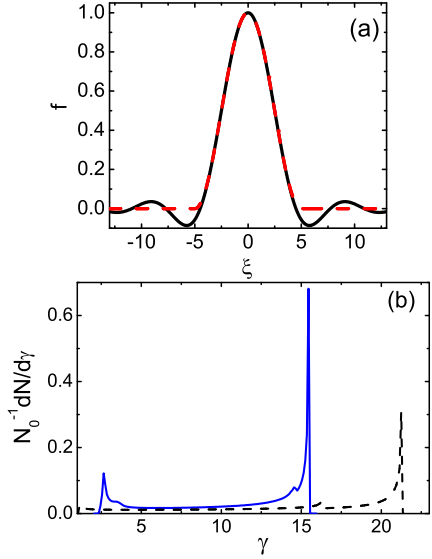}
\caption{(a) Focal-plane L-beam field profiles 
$f_{ult}(\xi)$ (solid) and $f_{trc} (\xi)$ (dashed).
(b) Distribution of transmitted electrons (N)
with the incident momentum $\rho_0=3$ over
relativistic factor $\gamma$  at the exit of the gate 
with the maximum field amplitude $f_{mx} = 12$
for the field profiles $f_{ult}$ (solid) and
$f_{cs}$ with $\xi_L=2$ (dashed).
$N_0$ -- the total number of incident electrons.}
\label{fig1n}
\end{center}
\end{figure}
Most substantial energy transfer is obtained at $w < \lambda $. 

Let us consider an example of the gate with
strongly relativistic maximum field amplitude, $ f_{mx} = 12 $. 
We study first the dependence  of averaged over the laser cycle
energy and momentum gains, i. e. 
$\langle \gamma_{out} \rangle - \gamma_0$ and 
$\langle \rho_{out} \rangle - \rho_0$ respectively,
$vs$ the incident momentum, $\rho_0$, see Fig. 2.
Fig. 2(a) shows that the energy gain, $\Delta \gamma \sim 6$ 
may far exceed the incident kinetic energy of E-beam. 
For the ultimate L-beam profile Eq. (4), even the 
phase averaged characteristics exhibit strong sensitivity 
to the incident electron momentum for $ \rho_0 \lesssim 1$, 
with electron energy significantly lower than the L-beam intensity.

In our simulations we considered electrons launched into 
the laser gate sufficiently far from its center, thus the
almost chaotic oscillations of $ \Delta \gamma $ 
are due to interference fringes (spatial tails) 
of the L-beam profile of Eq. (4).
The effect of these fringes is readily 
verified by considering a truncated 
L-beam profile without the fringes, 
$ f_{trc} (\xi)$, which in principle
can be realized in a spherical resonator of 
the radius $\xi_1$, where $ \xi_1 \approx 4.49 $ 
is the first zero of $ f_{ult}(\xi)$.
The field in this case is given by Eq. (4) for 
$ | \xi | \leq \xi_1 $, and zero otherwise.
Indeed, one can see that in this case there 
are no chaotic resonances in $ \Delta \gamma $, Fig. 2(b).  
In fact, the profile $ f_{trc} (\xi)$ can be 
approximated with a very high accuracy by 
\cite{cit2,cit2a}
\begin{equation}
f_{cs}(\xi)=f_{mx} \cos^2{( \xi/\xi_L)} \ \  {\rm at} \ \ | \xi/\xi_L | \leq \pi/2 
\ \ \ \
{\rm and} \ \ f_{cs} = 0 \ \  {\rm otherwise}
\label{eq5}
\end{equation}
with $ \xi_L \approx 3.18 $ (see Fig. 1(a)). 
The largest difference between $ f_{cs} $ and $ f_{trc} $ 
occurs near $ \xi_1 $, where $ f_{cs} ( \xi ) $ 
has a smoothly vanishing first derivative, 
in contrast to $ f_{trc} ( \xi ) $. 
Yet Fig. 2(a) shows that $\Delta \gamma$ 
for $f_{cs}$ and $f_{trc}$ nearly coincide at $ \rho_0 \gtrsim 2$. 
Even though the fringes of the $f_{ult}$ cause 
chaotic resonances at low incident electron energy,
the order of magnitude of the energy transfer for those 
L-beam profiles remains the same. 
Thus, for an approximate analysis one can work 
with the profile $ f_{cs} ( \xi )$, 
which is more convenient to use for numerical simulations, 
and, as we show below, allows for rigorous analytical treatment.
Another advantage of $ f_{cs} ( \xi )$ is that 
it has a free controllable parameter $ \xi_L $, 
which is related to the spot-size by 
$ \xi_L = 4 w / \lambda $. 
At $ \xi_L \gg 1 $ the profile $ f_{cs} $ is 
closely mimicking the Gaussian beam \cite{cit2,cit2a}. 
Tighter field profiles due to the presence of 
non-radiative fields will be characterized by $ \xi_L < \xi_1$ 
(see e.g. Fig 2(a), $ \xi_L = 2 $).

\begin{figure}[h]
\begin{center}
\includegraphics[width=8.6cm]{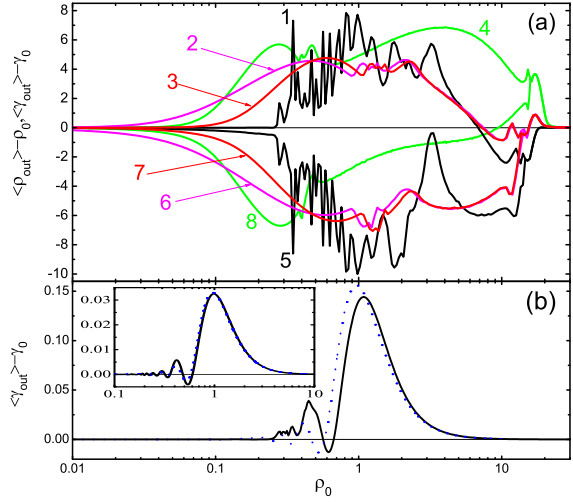}
\caption{Phase-averaged change of electron energy [curves 1-4 in (a), (b)], $\langle \Delta \gamma \rangle$  
and momentum (5-8), $\langle \Delta \rho \rangle$ \emph{vs} 
incident electron momentum, $\rho_0$. 
(a) Numerical simulation for a strongly relativistic laser gate ($f_{mx} = 12$) with the profile $f_{ult}$ (1,5); 
$f_{trc}$ (2,6); $f_{cs}$, $\xi_L=3.18$ (3,7); $f_{cs}$, $\xi_L=2.0$ (4,8).
(b) Numerical (solid) and analytical (Eq. (7), dotted line) results for $f_{cs}$, $\xi_L=2$, $f_{mx}=0.6$ (main plot) and $f_{mx}=0.23$ (inset).}
\label{fig2n}
\end{center}
\end{figure}

To further detail this energy transfer, 
next to the phase-averaged characteristics 
we consider another important feature:
the energy spectrum or distribution of scattered electrons
for the $fixed$  incident momentum; we consider
an example of $\rho_0=3$.
Because of the phase-dispersion, there are, in general, 
transmitted and reflected from the laser gate electrons. 
Fig. 1(b) depicts distribution $N_0 d N / d \gamma$,
(where $N_0$ is the total number of incident electrons)
of transmitted electrons ($N$)
over relativistic factor $\gamma$  at the exit of the gate
for two different
field profile models: $f_{ult} (\xi) $, Eq. (4), solid line,
and $f_{cs} (\xi)$ with $\xi_L=2$, Eq. (5), dashed line.
The incident relativistic factor here
is $\gamma_0 = {( 1 + \rho_0^2 )}^{1/2} \approx 3.16$,
which corresponds to the incident energy of electrons of $1.1$ MeV.
On the other hand, see Fig. 1(b),
the energy distribution of the transmitted electrons peaks
near much larger $ \gamma$, which, depending on the model,
is either $ \gamma \approx 15$ or $ \gamma \approx 22$,
which illustrates again a huge net transfer 
of energy from the laser to E-beam
with relativistic factor gain $\Delta \gamma \sim 12 - 19$,
or energy gain of $\sim 6.3$ MeV 
for the profile $f_{ult} (\xi) $, Eq. (4).

Remarkably, electrons acquired this energy gain
over the distance of $\sim \lambda$. 
This corresponds to a huge accelerating 
gradient $\sim 52$ GeV/cm, i.e. about 
two orders of magnitude higher than the 
acceleration gradient facilitated by the plasma-wakefield accelerator 
\cite{cit3,cit3a,cit3b,cit4}. 
Respectively, higher accelerating 
gradient $\sim 0.1$ TeV/cm is attained for 
$ \xi_L = 2 $ (Fig. 1(b)).
The reflected electrons acquire a similar 
energy gain and their spectrum has also a well pronounced peak.
Since in our geometry the L-beam and E-beam are orthogonal to each other, one can envision cascade design of the accelerating elements. 

In the above calculations, electrons were launched in the focal anti-node plane.
Any spatial deviation of electrons from that plane, inevitable in realistic situations, would result in transverse components of the Lorentz 
force coming into play because of interaction of electrons with all other, besides $E_x$, 
components of electric, as well as non-vanishing $\vect{H}$-field. However, our numerical simulations, based on exact 
solution for EM-fields \cite{cit6} producing the profile (4), showed that these effects are not drastic.  
For instance, for $f_{mx} = 12$, the angular divergence of the transmitted electrons, assumed to be parallel at input with $\rho_0 = 3$,
has $\theta_{HM} \approx 5$ and $23$ mrad for the incident E-beam spot size of $0.0025 \lambda$ and $0.025 \lambda$ respectively,
where $\theta_{HM}$ is the (scattering) angle between output and input electron momenta, taken at the half-maximum of the 
transmitted electron angular distribution. 

While profile of Eq. (4) reflects the limiting focusing 
of free-space radiative fields, it is not a general physical limit: 
using non-radiative, evanescent field in the material structures, one can engineer \emph{arbitrary tight} EM-profiles.
In fact, finer-than-$\lambda$ distributions have been successfully attained in microwave klystrons 
by using Hansen resonator \cite{cit8,cit8a,cit8b} with sub-$\lambda$ spacing between the field confining metallic grids 
\cite{cit9}.
Another way to attain sub-$\lambda$ field profiles is to make 
electrons run through the well known diffraction 
field pattern \cite{cit10} near a straight edge. 
Propagation of electrons in the sub-$\lambda$ confined relativistic fields brings about 
a ``quasi-static'' behavior, whereby the L-field can be assumed almost ``frozen'' in time as an electron passes through it.
In this case, instead of the amplitude of the field, the defining parameter is a full field area along 
the ``line-of-fire'', $A_L = \int_{-\infty}^{\infty}f(\xi)d\xi$, and 
\begin{equation}
\langle \Delta \gamma \rangle = \pi^{-1} [ A_L^2-(\gamma_0-1)^2 ]^{1/2}\ \ \ {\rm if}\ \ A_L>\gamma_0-1,
\label{eq6}
\end{equation}
and $\langle \Delta \gamma \rangle = 0$ otherwise. Eq. (6) predicts the maximum energy gain $\langle \Delta \gamma_{mx} \rangle = A_L / \pi$.

To relate the above results to more common and well analyzable situations of non-relativistic E-beams or low-gradient
L-beams, with a low inelastic scattering, we consider the electron momentum
dominance case $\rho_0 \gg f_{mx} $ \cite{cit2,cit2a}. 
Here, the E-beam runs \emph{via} the gate almost unhindered,
which allows for a Born-like approximation by assuming the laser field as a perturbation.
The zero-order approximation is an undisturbed electron motion, $\rho^{(0)} = \rho_0 $ and $\xi^{(0)} =-\pi\xi_L/2+\beta_0\tau$, with 
$\tau=0$ corresponding to the entering point. We look for the solution 
as a series \cite{cit11,cit11a}: 
$s=s^{(0)}+\Delta s^{(1)}+ \Delta s^{(2)} + ...$, 
where $s(\tau)$ is a generic variable of the problem, 
with $s^{(n)}={\rm O} [(f_{mx}/\rho_0)^n]$.

Evaluating 1-st order corrections at the exit, 
$\Delta s_{out}^{(1)} $, we find that for some  phases $\phi$ there 
is momentum and energy gain, whereas for other there is a loss of both. 
If again the electrons arrive randomly, we have to average $\Delta s_{out}$ over the ensemble, i.e. over the phase $0 < \phi \leq 2\pi$,
and arrive as expected, at a zero gain/loss: $\langle \Delta \rho^{(1)}\rangle = \langle \Delta \xi^{(1)}\rangle = 0$.
Thus, one has to use a 2-nd order approximation, 
$\Delta s^{(2)}$(which is typical for Cherenkov 
and transition radiation lasers  
\cite{cit12,cit12a,cit12b} and NLO of single electron  \cite{cit11,cit11a}), 
to account for a small spatial electron displacement 
affecting the force seen by an electron.
Our calculations yield:
\begin{equation}
\frac{\langle\Delta \gamma^{(2)}\rangle}{f_{mx}^2}= \frac{B^3 \sin{(\pi/B)}}{2 \gamma_0^3(1-B^2)^2} \times
\left [ \frac{B(3-B^2)}{(1- B^2)} \sin \left( \frac{\pi}{B} \right )- \pi\cos \left( \frac{\pi}{B} \right) \right],
\label{eq7}
\end{equation}
where $B=2\beta_0/\xi_L$.  
Fig. 2(b) depicts analytical (7) and numerical results with $\xi_L=2$
when $f_{mx}=0.23$ (inset), where they are almost identical, and when $f_{mx}=0.6$, which, at the gain $\sim 15$\% still
shows a good agreement between those two. 
For low field gradient, $\xi_L \gg 1$, we have $B \ll 1$ when $\beta_0 \sim 1$, and the envelope of oscillations in (7) is 
\begin{equation}
\langle \Delta \gamma^{(2)} \rangle_{env} \approx 2 \pi f_{mx}^2(\rho_0/\xi_L\gamma_0^2)^3 
\label{eq8}
\end{equation}
with the peak parameters 
$(\rho_0)_{pk}=1$, and 
\begin{equation}
\langle \Delta \gamma^{(2)} \rangle_{pk} =\pi f_{mx}^2/(4\xi_L^3), 
\label{eq9}
\end{equation}
so the larger the $\xi_L $, 
the smaller energy transfer, as expected. 
With $\rho_0$ increasing beyond the $(\rho_0)_{pk}$, the inelastic effect
rapidly vanishes as $\sim \gamma_0^{-3}$. 
It is worth noting that at certain intervals in Fig. 2(b), 
there is an energy loss by the E-beam, 
$\langle \Delta \gamma^{(2)}\rangle < 0$: 
the energy is transferred from the E-beam to the L-beam, 
resulting in \emph{stimulated emission}, 
which can in principle be used for free-electron lasing.

\section{Klystron-like temporal focusing of electron beam}
The ``laser as perturbation'' approach is instrumental 
also in the analysis of another feature: 
dramatic klystron-like \cite{cit8,cit8a,cit8b}
temporal focusing (E-bunch formation) of the E-beam after 
it passes through the laser gate.
By considering \emph{each individual phase} 
before the averaging, our results show a substantial momentum/speed modulation,
so that the exiting E-beam bears the 
memory of the interaction within \emph{each of laser cycles}.
While one of the contributing factors is an accumulated extra time delay, $\Delta \tau_{out}^{(1)}(\phi)$, 
the defining contribution to the bunching effect is an accumulated extra momentum at the exit:
\begin{equation}
\Delta \rho_{out}^{(1)}(\phi)=-\frac{f_{mx}B^2}{1-B^2} \sin \left( \frac{\pi}{B} +\phi \right) 
\sin \left( \frac{\pi}{B} \right).
\label{eq10}
\end{equation}
Similarly to the thin-lens approximation in optics, we assume here that all the electrons need approximately the same time to get to the exit, 
i. e. output face of the laser gate, but accrue different velocities/momenta.
Counting now the distance run by each electron from the output face, $\xi(\tau=\phi) = 0$, we can construct their ``beyond the gate'' 
time-lines by evaluating the time, $\tau_\xi$, counted from e.g. beginning of the laser cycle, as laser phase $\phi$ plus
the time lapsed from the moment an electron left that exit face, to the moment it reached distance $\xi$
\begin{equation}
\tau_\xi(\phi) \approx \phi + 
\frac{\xi}{\beta_0} \left[1-\frac{\Delta \rho_{out}^{(1)}(\phi)}{\beta_0 \gamma_0^3} \right]
\label{eq11}
\end{equation}
where $\Delta \beta_{out}(\phi)\approx \Delta \rho_{out}^{(1)} (\phi)/\gamma_0^3 $ and $\Delta \rho_{out}^{(1)}$ is determined by (10). 
The current density temporal profile, $j_\xi(\tau)$, of E-beam at a distance $\xi$ can in general be written as
$j_\xi(\tau)/j_0=\left[ 2 \pi (d \tau/d \phi)\right]^{-1} (\geq 0)$, where $j_0=j_{\xi=0}$. 
In approximation (7) we have:
\begin{equation}
2 \pi j_\xi(\tau)/j_0=\left[1-(\xi/\xi_f) \cos{(\pi/B+\phi)} \right]^{-1},
\label{eq12}
\end{equation}
where $\xi_f > 0$ is the focal distance,
\begin{equation}
\xi_f=\left| \frac{\pi(\gamma_0 \xi_L )^3 (B^2-1)}{8 A_L \sin{(\pi/B)}} \right|
\label{eq13}
\end{equation}
Eq. (12) is true only when $\xi \leq \xi_f $. 
As the E-beam approaches the focusing point, $\xi \rightarrow  \xi_f$, so that $\Delta \xi = \xi_f - \xi \ll \xi_f $,
(12) describes a Lorentzian pulse at $\xi$, near the moment  $(\tau_\xi)_{pk}-\xi/\beta_0 = \phi_{pk} =- \pi/B$
\begin{equation}
j_\xi(\tau) \approx \frac{j_0}{2\pi} \frac{1/\delta}{1 +(\Delta \tau)^2/2 \delta^3} \ \ \  {\rm where} \ \ \
\delta = \frac{\Delta \xi}{\xi_f}.
\label{eq14}
\end{equation}
\begin{figure}
\begin{center}
\includegraphics[width=8.6cm]{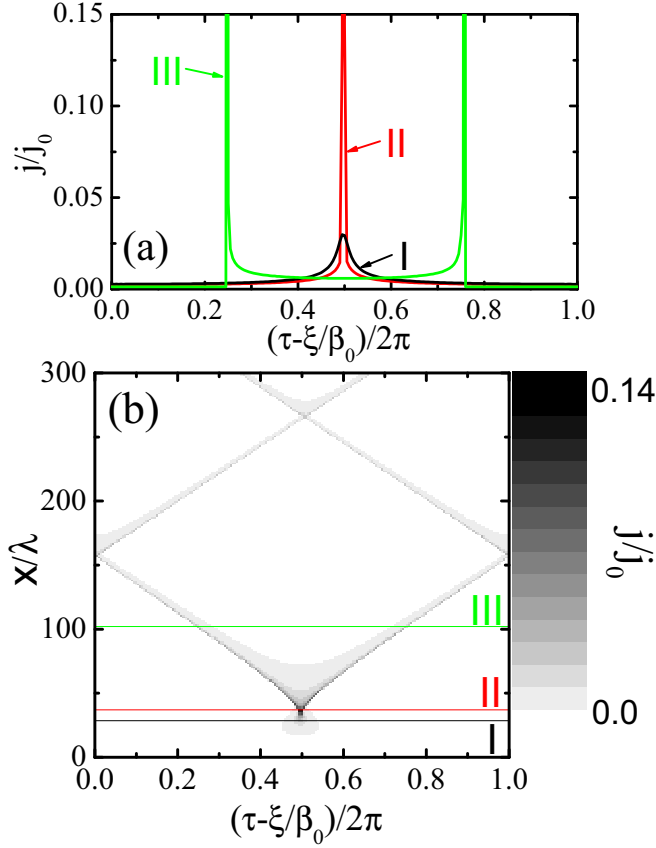}
\caption{Focusing/bunching of electrons ($\rho_0 = 1$) 
by the laser gate ($\xi_L = \pi $, 
$A_L = 0.2$). Spatio-temporal (b) and temporal (a) at $x/\lambda = 28.5$ (I), $ 36.5$ (II), and $102$ (III) profiles of the current density, $j$.}
\label{fig3n}
\end{center}
\end{figure}

The diverging focusing remains even if one accounts for the time delay at the exit face.
The divergence at $\xi \rightarrow \xi_f $ is removed and the peak resolved by accounting for
other physical factors, e.g. non-monoenergetic electrons, the Coulomb repulsion at the focusing point, 
finite width of E-beam due to laser magnetic field, etc. 
The ultimate limitation is imposed by uncertainty principle, due to which the finite bunch length is:
\begin{equation}
\Delta \tau_f = \omega \Delta t_f = (\hbar \omega / mc^2 )/ \Delta \gamma_{max} 
\label{eq15}
\end{equation}
where $\Delta \gamma_{max} $ is the maximum modulation of the energy of the output E-beam due to laser driving.
This limitation is similar to diffraction limit of focusing in optics.
Using (10) for $\Delta \rho_{out} $ with $B=2$ and having in mind that $\Delta \gamma \approx \beta_0 \Delta \rho $,
we have 
\begin{equation}
\Delta \gamma_{max} = 8 A_L /(3 \pi ). 
\label{eq16}
\end{equation}
If $A_L \sim \Delta \gamma_{max} = 0.2$, and $ \hbar \omega \sim 1$ eV,
we have $\Delta \tau_f \sim 10^{-5} $, which corresponds to 
$\Delta t_f \sim 30$ zs, where $1$ zs $= 10^{-21} $ s  
(zepto-second \cite{cit13}).
An estimate based on the spread of E-beam energy spectrum  
$\sim \Delta \gamma_{in}\sim 10^{-6}$ 
as in an electron microscope \cite{cit14}, $B=2$ 
and  $A_L \sim 0.2$, yields $\Delta \tau_f \sim 1.5 \times 10^{-5}$,
or $\Delta t_f \sim 45$ zs, which comes close to the quantum limit.

Equations (8)-(10) remain valid only if the time-lines $\xi (\tau)$ of individual electrons do not cross each other, i.e. until 
crossing of many time-lines at  $ \xi = \xi_f $.
Beyond that point, faster electrons over-run the slower ones, 
forming two shock waves propagating in opposite directions and making a 
``two-horn'' temporal profile at each point 
$\xi$, with each horn corresponding to the 
divergence in the density current, see Fig. 3(a);
Fig. 3(b) depicts spatiotemporal and temporal profiles
of the current density, $j$ .
These shock-waves are typical for a gas of 
weakly-interacting particles, 
whereby the slower particles running in front 
of the group at some point (focus) 
are overrun by the faster particles coming from behind 
and creating for a moment infinite density at certain (focal) point.
Aside from the direct analogy to the bunch formation 
in \emph {mw} klystron \cite{cit8,cit8a,cit8b}, 
this process is similar to spatial ray focusing 
and caustics in geometric optics \cite{cit10}, 
as well as shock waves in 
astrophysics \cite{cit15} and in Coulomb explosions \cite{cit16}.

\section{Conclusion}
In conclusion, we demonstrated the feasibility of multi-MeV electron acceleration corresponding to $\sim 0.1$ TeV/cm 
acceleration gradients, based upon strong inelastic 
scattering of electrons in an ultra-gradient 
relativistically-intense laser field.
We also predicted the formation of zepto-second electron 
bunches due to strong after-scattering electron beam modulation.

This work was supported by AFOSR. 


\begin{thebibliography}{99}
\bibitem{cit1} H. A. H. Boot and R. B. R-S. Harvie, 
"Charged particles in a non-uniform radio-frequency field,"
Nature {\bf 180}, 1187-1187 (1957)
\bibitem{cit1a} A. V. Gaponov and M.A. Miller, 
"Potential wells for charged particles in a
high-frequency electromagnetic field," 
Sov. Phys. JETP {\bf 7}, 168-169 (1958)
\bibitem{cit1b} T. W. B. Kibble, 
"Refraction of electron beams by intense electromagnetic waves,"
Phys. Rev. Lett. {\bf 16}, 1054-1056 (1966)
\bibitem{cit1c} M. V. Fedorov, 
"Stimulated scattering of electrons by photons and adiabatic
switching on hypothesis," Opt. Commun. {\bf 12}, 205-209 (1974)
\bibitem{cit2} A. E. Kaplan and A. L. Pokrovsky, 
"Fully relativistic theory of the ponderomotive force 
in an ultraintense standing wave," Phys. Rev. Lett. {\bf 95}, 
053601(1-4) (2005)
\bibitem{cit2a}
A. L. Pokrovsky and A. E. Kaplan, 
"Relativistic reversal of the ponderomotive force in a standing laser wave," 
Phys. Rev. A {\bf 72}, 043401(1-12) (2005)
\bibitem{cit3} C. Gahn, G. D. Tsakiris, A. Pukhov, J. Meyer-ter-Vehn, 
G. Pretzler, P. Thirolf, D. Habs, K. J. Witte, 
"Multi-MeV electron beam generation by direct 
laser acceleration in high-density plasma channels," 
Phys. Rev. Lett. {\bf 83}, 4772-4775 (1999)
\bibitem{cit3a} 
C. G. R. Geddes, C. Toth C, J. van Tilborg, 
E. Esarey, C. B. Schroeder, D. Bruhwiler, C. Nieter, J. Cary, 
W. P. Leemans, "High-quality electron beams from a 
laser wakefield accelerator using plasma-channel guiding," 
Nature {\bf 431}, 538-541 (2004)
\bibitem{cit3b}
J. Faure, Y. Glinec, A. Pukhov, S. Kiselev, S. Gordienko, 
E. Lefebvre, J. P. Rousseau, F. Burgy, V. Malka, 
"A laser-plasma accelerator producing 
monoenergetic electron beams," Nature {\bf 431}, 541-544 (2004)
\bibitem{cit4} M. J. Hogan, C. D. Barnes, C. F. Clayton, 
"Multi-GeV energy gain in a plasma-wakefield accelerator," 
Phys. Rev. Lett. {\bf 95}, 054802(1-4) (2005). 
\bibitem{cit5} G. Shvets, "Beat-Wave Excitation 
of Plasma Waves Based on Relativistic Bistability," 
Phys. Rev. Lett. {\bf 93}, 195004(1-4) (2004)
\bibitem{cit6}
We neglect here the "radiation friction" force on electron;
this was supported by all our estimates
and numerical simulations for the specific situation.
The time for an electron to pass through
the laser gate is very short, and
for the radiation friction to affect the motion,
one needs $\gamma\sim 10^{2} - 10^{3}$,
which is beyond the domain of interest.
Also, when addressing the EM-electron interaction,
we use classical approach, since in the cases
of interest, a typical number of photons absorbed by an electron per pass,
is of the order of $m c^{2} / \hbar \omega \sim 10^{6}$.
\bibitem{cit7} L. D. Landau, E. M. Lifshitz, and L. P. Pitaevskii, 
\emph {Electrodynamics of Continuous Media}, p. 312 (Pergamon, New-York, 1984)
\bibitem{cit8} R. H. Varian and S. F. Varian,
"A High Frequency Oscillator and Amplifier," 
J. Appl. Phys. {\bf 10}, 321-327 (1939)
\bibitem{cit8a} 
D. L. Webster, "Cathode-Ray Bunching," J. Appl. Phys. {\bf 10}, 
501-508 (1939)
\bibitem{cit8b}
W. W. Hansen, "A Type of Electrical Resonator," J. Appl. Phys.
{\bf 9}, 654-663 (1938)
\bibitem{cit8c}
W. W. Hansen and  R. D. Richtmyer, 
"On Resonators Suitable for Klystron Oscillators," 
J. Appl. Phys.  {\bf 10}, 189-199 (1939).
\bibitem{cit9} It would be a challenging but greatly rewarding endeavor to develop Hansen-like resonators for optical domain; 
there is no physical restriction on the size of the field inhomogeneity $\xi_L$.
\bibitem{cit10} M. Born and E. Wolf, 
\emph{Principles of Optics}, Pergamon Press, 6th Ed. 1980, p. 127.
\bibitem{cit11} A. E. Kaplan, 
``Relativistic nonlinear optics of a single cyclotron electron," 
Phys. Rev. Lett. {\bf 56}, 456-459 (1986)
\bibitem{cit11a} A. E. Kaplan, Y. J. Ding, 
``Hysteretic and multiphoton optical resonances 
of a single cyclotron electron," IEEE J. Quantum Electron. {\bf 24}, 
1470-1482 (1988)
\bibitem{cit12} W. Becker and J. K. McIver, Phys. Rev. A {\bf 31}, 
783-789 (1985)
\bibitem{cit12a} A. E. Kaplan and S. Datta, ``Extreme-ultraviolet and X-ray 
Emission and Amplification by Non-relativistic Beams Traversing a Superlattice," Appl. Phys. Lett. {\bf 44}, 661-663 (1984)
\bibitem{cit12b}
S. Datta and A. E. Kaplan, ``Quantum Theory of Spontaneous and Stimulated Resonant Transition Radiation,"
 Phys. Rev. A. {\bf 31}, 790-796 (1985).
\bibitem{cit13}  A. E. Kaplan and P. L. Shkolnikov, 
``Lasetron: a proposed source of powerful 
nuclear-time-scale electromagnetic bursts," 
Phys. Rev. Lett. {\bf 88}, 074801(1-4) (2002).
\bibitem{cit14} V. Ravikumar, R. P. Rodrigues, V. P. Dravid,
"Space-charge distribution across internal interfaces 
in electroceramics using electron holography,"
J. Am. Ceram. Soc. {\bf 80}, 1117-1130 (1997).
\bibitem{cit15} Ya. B. Zel'dovich and I. D. Novikov, 
\emph{Relativistic Astrophysics}, v. 2: 
\emph{The structure and Evolution of the Universe}, 
p. 361 (The Univ. Chicago Press, Chicago, 1983).
\bibitem{cit16} A. E. Kaplan, B. Y. Dubetsky, and P. L. Shkolnikov, 
"Shock-shells in Coulomb explosion of nanoclusters," 
Phys. Rev. Lett. {\bf 91}, 143401(1-4) (2003).
\end{thebibliography}
\end{document}